\documentclass[11pt]{article}
\usepackage{amssymb,latexsym,amsmath,amsbsy}
\usepackage[dvips]{graphicx}
\usepackage{cite}
\newtheorem{theo}{Theorem}

\newtheorem{coro}[theo]{Corollary}

\newtheorem{prop}[theo]{Proposition}
\def\nn{\nonumber}

\def\ch{\mathop{\rm char}\nolimits}

\def\qdots{\mathinner{\mkern1mu\raise1pt\vbox{\kern7pt\hbox{.}}\mkern2mu
 \raise4pt\hbox{.}\mkern2mu\raise7pt\hbox{.}\mkern1mu}}
\def\Z{{\mathbb Z}}

\def\R{{\mathbb R}}
\def\C{{\mathbb C}}
\def\gl{\mathfrak{gl}}

\def\u{\mathfrak{u}}
\def\h{\mathfrak{h}}

\def\sp{\mathfrak{sp}}
\def\osp{\mathfrak{osp}}

\renewcommand{\theequation}{\arabic{section}.{\arabic{equation}}}

\begin{document}
\begin{center}
{\Large \bf
The paraboson Fock space and unitary irreducible representations \\[2mm]
of the Lie superalgebra $\osp(1|2n)$ }\\[5mm]
{\bf S.~Lievens\footnote{E-mail: Stijn.Lievens@UGent.be}, }
{\bf N.I.~Stoilova}\footnote{E-mail: Neli.Stoilova@UGent.be; Permanent address:
Institute for Nuclear Research and Nuclear Energy, Boul.\ Tsarigradsko Chaussee 72,
1784 Sofia, Bulgaria} {\bf and J.\ Van der Jeugt}\footnote{E-mail:
Joris.VanderJeugt@UGent.be}\\[1mm]
Department of Applied Mathematics and Computer Science,
Ghent University,\\
Krijgslaan 281-S9, B-9000 Gent, Belgium.
\end{center}

\vskip 2cm
\addtolength{\parskip}{2mm}

\begin{abstract}
It is known that the defining relations of the orthosymplectic Lie superalgebra 
$\osp(1|2n)$ are equivalent to the defining (triple) relations of $n$ pairs of paraboson operators $b^\pm_i$.
In particular, with the usual star conditions, this implies that the ``parabosons of order $p$''
correspond to a unitary irreducible (infinite-dimensional) lowest weight representation $V(p)$ of $\osp(1|2n)$. 
Apart from the simple cases $p=1$ or $n=1$, these representations
had never been constructed due to computational difficulties, despite their importance.
In the present paper we give an explicit and elegant construction of these representations $V(p)$,
and we present explicit actions or matrix elements of the $\osp(1|2n)$ generators.
The orthogonal basis vectors of $V(p)$ are written in terms of Gelfand-Zetlin patterns,
where the subalgebra $\u(n)$ of $\osp(1|2n)$ plays a crucial role.
Our results also lead to character formulas for these infinite-dimensional $\osp(1|2n)$ representations.
Furthermore, by considering the branching $\osp(1|2n) \supset \sp(2n) \supset \u(n)$, 
we find explicit infinite-dimensional unitary irreducible lowest weight representations
of $\sp(2n)$ and their characters.
\end{abstract}


\setcounter{equation}{0}
\section{Introduction} \label{sec:Introduction}%

The classical notion of Bose operators or bosons has been generalized a long time ago
to parabose operators or parabosons~\cite{Green}. 
These parabosons are of interest in many applications, in particular in quantum field 
theory~\cite{Haag, Ohnuki, Wu},
generalizations of quantum statistics (para-statistics)~\cite{Green, Greenberg, Nelson, Macrae}, and in Wigner 
quantum systems~\cite{Wigner, Palev86, Palev02, Lievens}.
The generalization of the usual boson Fock space is characterized by a parameter $p$,
referred to as the order. For a single paraboson, $n=1$, the structure of the paraboson
Fock space is well known~\cite{Palev1}. 
Surprisingly, for a system of $n$ parabosons with $n>1$, the structure of the paraboson
Fock space is not known, even though it can in principle be constructed by means of 
the so-called Green ansatz~\cite{Green}. The computational difficulties of the Green ansatz are
related to finding a proper basis of an irreducible constituent of a $p$-fold
tensor product~\cite{Greenberg}.

An important result was given by Ganchev and Palev~\cite{Ganchev}, who observed that the 
triple relations for $n$ pairs of parabosons are in fact defining relations 
for the orthosymplectic Lie superalgebra $\osp(1|2n)$~\cite{Kac}.
This implies also that the paraboson Fock space of order $p$ is a certain
infinite-dimensional unitary irreducible representation (unirrep) of $\osp(1|2n)$.
However, the construction of this representation, for $n>1$ and arbitrary $p$,
also turned out to be difficult. 
In the present paper we present a solution to this problem. 
Our solution is based upon group theoretical techniques, in particular
related to the branching $\osp(1|2n)\supset \sp(2n)\supset \u(n)$. 
This allows us to construct a proper Gelfand-Zetlin (GZ) basis for 
some induced representation~\cite{Kac1}, from which the basis for the irreducible
representation follows. 
Our result is thus a complete solution to the problem: for the 
paraboson Fock space we give not only an orthogonal GZ-basis but also
the action (matrix elements) of the paraboson operators in this basis.

The structure of the paper is as follows. In section~2, we recall the definition of
parabosons and of the paraboson Fock space $V(p)$. In section~3, we discuss the important
relation between parabosons and the Lie superalgebra $\osp(1|2n)$, and give a 
description of $V(p)$ in terms of representations of $\osp(1|2n)$. The following section
is devoted to the first non-trivial example of $\osp(1|4)$. The analysis of the
representations $V(p)$ for $\osp(1|4)$ is performed in detail, and the techniques
used here can be lifted to the general case $\osp(1|2n)$. This general case is
investigated in section~5, where the main computational result (matrix elements) is
given in Proposition~\ref{prop-main} and the main structural result (characters) in
Theorem~\ref{theo-main}. Section~6 is devoted to the branching to the even subalgebra
$\sp(2n)=\sp(2n,\R)$, and gives some $\sp(2n)$ characters. We end the paper with some final remarks.

\setcounter{equation}{0}
\section{The paraboson Fock space $V(p)$} \label{sec:Fock}

Before introducing the paraboson Fock space, let us recall some aspects of the
usual boson Fock space.
For a single pair ($n=1$) of boson operators $B^+$, $B^-$, the defining relation
is given by
\begin{equation}
 [B^-,B^+]=1 .
 \label{boson}
\end{equation}
The boson Fock space is defined as a Hilbert space with vacuum vector $|0\rangle$, in
which the action of the operators $B^+$, $B^-$ is defined and satisfies
\begin{equation}
\langle 0|0\rangle=1, \qquad B^- |0\rangle = 0, \qquad (B^\pm)^\dagger = B^\mp.
\end{equation}
Moreover, under the action of the algebra
spanned by $\{B^+, B^-,1\}$ (subject to~\eqref{boson}), the Hilbert space is irreducible.
A set of basis vectors of this space, denoted by $V(1)$, is given by
\begin{equation}
 |k\rangle = \frac{(B^+)^k}{\sqrt{k!}}|0\rangle,\qquad k\in\Z_+=\{0,1,2\ldots\}.
 \label{basis-boson}
\end{equation}
These vectors are orthogonal and normalized. 
The space $V(1)$ is a unitary irreducible representation (unirrep) of the Lie superalgebra
$\osp(1|2)$~\cite{Palev1} (see next paragraph). 
It is also the direct sum of two unirreps of the Lie algebra $\sp(2)=\sp(2,\R)$
(one with even $k$'s and one with odd $k$'s), known as the metaplectic representations
(certain positive discrete series representations of $\sp(2)$)~\cite{B,I,S}.

For a single pair ($n=1$) of paraboson operators $b^+$, $b^-$~\cite{Green}, the defining relation
is a triple relation (with anticommutator $\{.,.\}$ and commutator $[.,.]$) given by
\begin{equation}
 [\{b^-,b^+\},b^\pm]=\pm 2 b^\pm .
 \label{paraboson}
\end{equation}
The paraboson Fock space~\cite{Palev1} is again a Hilbert space with vacuum vector $|0\rangle$, 
defined by means of
\begin{align}
& \langle 0|0\rangle=1, \qquad b^- |0\rangle = 0, \qquad (b^\pm)^\dagger = b^\mp,\nn\\
& \{b^-,b^+\} |0\rangle = p\, |0\rangle,
\end{align}
and by irreducibility under the action of $b^+$, $b^-$.
Herein, $p$ is a parameter, known as the order of the paraboson.
In order to have a genuine inner product, $p$ should be positive and real: $p>0$.
A set of basis vectors for this space, denoted by $V(p)$, is given by
\begin{equation}
 |2k\rangle = \frac{(b^+)^{2k}}{2^k\sqrt{k!(p/2)_k}}|0\rangle,\qquad
 |2k+1\rangle = \frac{(b^+)^{2k+1}}{2^k\sqrt{k!2(p/2)_{k+1}}}|0\rangle.
 \label{basis-paraboson}
\end{equation}
This basis is orthogonal and normalized; the symbol $(a)_k = a(a+1)\cdots (a+k-1)$ 
is the common Pochhammer symbol.

If one considers $b^+$ and $b^-$ as odd generators of a Lie superalgebra,
then the elements $\{b^+,b^+\}$, $\{b^+,b^-\}$ and $\{b^-,b^-\}$ form a basis
for the even part of this superalgebra. Using the relations~\eqref{paraboson} it
is easy to see that this superalgebra is the orthosymplectic Lie superalgebra
$\osp(1|2)$, with even part $\sp(2)$. 
The paraboson Fock space $V(p)$ is then a unirrep of $\osp(1|2)$. It splits as
the direct sum of two positive discrete series representations of $\sp(2)$:
one with lowest weight vector $|0\rangle$ (lowest weight $p/2$) and basis 
vectors $|2k\rangle$, and one with lowest weight vector $|1\rangle$ (lowest weight
$1+p/2$) and basis vectors $|2k+1\rangle$. For $p=1$ the paraboson Fock space
coincides with the ordinary boson Fock space. This also follows from the general action
\begin{equation}
(b^-b^+-b^+b^-)|2k\rangle = p\,|2k\rangle,\qquad
(b^-b^+-b^+b^-)|2k+1\rangle = (2-p)\,|2k+1\rangle.
\end{equation}

Let us now consider the case of $n$ pairs of boson operators $B_i^\pm$ ($i=1,2,\ldots,n$),
satisfying the standard commutation relations
\begin{equation}
 [B_i^-,B_j^+]=\delta_{ij}.
 \label{n-boson}
\end{equation}
The $n$-boson Fock space is again defined as a Hilbert space with vacuum vector $|0\rangle$, 
with
\begin{equation}
\langle 0|0\rangle=1, \qquad B_i^- |0\rangle = 0, \qquad (B_i^\pm)^\dagger = B_i^\mp
\qquad (i=1,\ldots,n).
\end{equation}
The Hilbert space is irreducible under the action of the algebra
spanned by the elements $1, B_i^+, B_i^-$ ($i=1,\ldots,n$), subject to~\eqref{n-boson}.
A set of (orthogonal and normalized) basis vectors of this space is given by
\begin{equation}
 |k_1,\ldots,k_n\rangle = \frac{(B_1^+)^{k_1}\cdots(B_n^+)^{k_n} }{\sqrt{k_1!\cdots k_n!}}|0\rangle,
 \qquad k_1,\ldots,k_n\in\Z_+ .
 \label{basis-n-boson}
\end{equation}
We shall see that this Fock space is a certain unirrep of the Lie superalgebra
$\osp(1|2n)$, with lowest weight $(\frac{1}{2},\ldots,\frac{1}{2})$.

We are primarily interested in a system of $n$ pairs of paraboson operators 
$b_j^\pm$ $(j=1,\ldots,n)$. The defining triple relations for such a system
are given by~\cite{Green}
\begin{equation}
 [\{ b_{ j}^{\xi}, b_{ k}^{\eta}\} , b_{l}^{\epsilon}]= (\epsilon -\xi) \delta_{jl} b_{k}^{\eta} 
 +  (\epsilon -\eta) \delta_{kl}b_{j}^{\xi}, 
 \label{n-paraboson}
\end{equation}
where $j,k,l\in\{1,2,\ldots,n\}$ and $\eta, \epsilon, \xi \in\{+,-\}$ (to be interpreted as $+1$ and $-1$
in the algebraic expressions $\epsilon -\xi$ and $\epsilon -\eta$).
The paraboson Fock space $V(p)$ is the Hilbert space with vacuum vector $|0\rangle$, 
defined by means of ($j,k=1,2,\ldots,n$)
\begin{align}
& \langle 0|0\rangle=1, \qquad b_j^- |0\rangle = 0, \qquad (b_j^\pm)^\dagger = b_j^\mp,\nn\\
& \{b_j^-,b_k^+\} |0\rangle = p\delta_{jk}\, |0\rangle,
\end{align}
and by irreducibility under the action of the algebra spanned by
the elements $b_j^+$, $b_j^-$ ($j=1,\ldots,n$), subject
to~\eqref{n-paraboson}. The parameter $p$ is referred to as the order of the paraboson system.
In general $p$ is thought of as a positive integer, and for $p=1$ the paraboson Fock
space $V(p)$ coincides with the boson Fock space $V(1)$. 
We shall see that also certain non-integer $p$-values are allowed.

Constructing a basis for the Fock space $V(p)$ turns out to be a difficult problem,
unsolved so far. 
Even the simpler question of finding the structure of $V(p)$ (weight structure)
is not solved. 
In the present paper we shall unravel the structure of $V(p)$, determine for which $p$-values
$V(p)$ is actually a Hilbert space, construct an orthogonal (normalized) basis for $V(p)$,
and give the actions of the generators $b_j^\pm$ on the basis vectors.

\setcounter{equation}{0}
\section{The Lie superalgebra $\osp(1|2n)$} \label{sec:osp12n}

The Lie superalgebra $\osp(1|2n)$~\cite{Kac} consists of matrices of the form
\begin{equation}
\left(\begin{array}{ccc} 
0&a&a_1\\
a_1^t&b&c\\
-a^t&d&-b^t
\end{array}\right),
\label{osp12n}
\end{equation}
where $a$ and $a_1$ are $(1\times n)$-matrices, $b$ is any $(n\times n)$-matrix,
and $c$ and $d$ are symmetric $(n\times n)$-matrices. The even elements have 
$a=a_1=0$ and the odd elements are those with $b=c=d=0$. 
It will be convenient to have the row and column indices running from $0$ to $2n$ 
(instead of $1$ to $2n+1$), and to denote by $e_{ij}$ the matrix with zeros everywhere except
a $1$ on position $(i,j)$. Then the Cartan subalgebra $\h$ of $\osp(1|2n)$ is 
spanned by the diagonal elements
\begin{equation}
h_j = e_{jj}-e_{n+j,n+j} \qquad (j=1,\ldots,n).
\label{h_j}
\end{equation}
In terms of the dual basis $\delta_j$ of $\h^*$, 
the odd root vectors and corresponding roots of $\osp(1|2n)$ are given by:
\begin{align}
& e_{0,k}-e_{n+k,0}  \leftrightarrow -\delta_k, \qquad k=1,\ldots ,n, \nn\\
& e_{0,n+k}+e_{k,0}  \leftrightarrow \delta_k, \qquad k=1,\ldots ,n. \nn
\end{align}
The even roots and root vectors are
\begin{align}
& e_{j,k}-e_{n+k,n+j}  \leftrightarrow \delta_j -\delta_k,  
\qquad j\neq k=1,\ldots ,n,\nn\\
& e_{j,n+k}+e_{k,n+j}  \leftrightarrow  \delta_j +\delta_k, \qquad j\leq k=1,\ldots ,n,\nn\\
& e_{n+j,k}+e_{n+k,j}  \leftrightarrow -\delta_j -\delta_k, \qquad j\leq k=1,\ldots ,n. \nn
\end{align}
If we introduce the following multiples of the odd root vectors 
\begin{equation}
b_{k}^+= \sqrt{2}(e_{0, n+k}+e_{k,0}), \qquad
b_{k}^-= \sqrt{2}(e_{0, k}-e_{n+k,0}) \qquad (k=1,\ldots , n)
\label{b-as-e}
\end{equation}
then it is easy to verify that these operators satisfy the triple relations~\eqref{n-paraboson}.
Since all even root vectors can be obtained by anticommutators $\{b_j^{\xi}, b_{ k}^{\eta}\}$,
the following holds~\cite{Ganchev}
\begin{theo}[Ganchev and Palev]
As a Lie superalgebra defined by generators and relations, 
$\osp(1|2n)$ is generated by $2n$ odd elements $b_k^\pm$ subject to the following (paraboson) relations:
\begin{equation}
 [\{ b_{ j}^{\xi}, b_{ k}^{\eta}\} , b_{l}^{\epsilon}]= (\epsilon -\xi) \delta_{jl} b_{k}^{\eta} 
 +  (\epsilon -\eta) \delta_{kl}b_{j}^{\xi}.
\label{pboson}
\end{equation}
\end{theo}
The paraboson operators $b_j^+$ are the positive odd root vectors, and the
$b_j^-$ are the negative odd root vectors.

Recall that the paraboson Fock space $V(p)$ is characterized by ($j,k=1,\ldots,n$)
\begin{equation}
 (b_j^\pm)^\dagger = b_j^\mp, \qquad b_j^- |0\rangle = 0, \qquad
 \{b_j^-,b_k^+\} |0\rangle = p\,\delta_{jk}\, |0\rangle.
 \label{pFock}
\end{equation}
Furthermore, it is easy to verify that
\begin{equation}
\{b_j^-,b_j^+\}=2 h_j \qquad(j=1,\ldots,n).
\label{bbh}
\end{equation}
Hence we have the following:
\begin{coro}
The paraboson Fock space $V(p)$ is the unitary irreducible representation of
$\osp(1|2n)$ with lowest weight $(\frac{p}{2}, \frac{p}{2},\ldots, \frac{p}{2})$.
\end{coro} 

In order to construct the representation $V(p)$~\cite{Palev4} one can use an induced module construction.
The relevant subalgebras of $\osp(1|2n)$ are easy to describe by means of the
odd generators $b_j^\pm$. 
\begin{prop}
A basis for the even subalgebra $\sp(2n)$ of $\osp(1|2n)$ is given by the elements
\begin{equation}
\{b_j^\pm, b_k^\pm\}\quad (1\leq j\leq k\leq n),\quad \{b_j^+, b_k^-\} \quad(1\leq j,k\leq n).
\label{sp2n}
\end{equation}
The $n^2$ elements 
\begin{equation}
\{b_j^+, b_k^-\} \qquad(j,k=1,\ldots,n)
\label{un}
\end{equation} 
are a basis for the $\sp(2n)$ subalgebra $\u(n)$.
\end{prop}

Note that with $\{b_j^+,b_k^-\}=2E_{jk}$, the triple relations~\eqref{pboson} imply the
relations $[E_{ij},E_{kl}]=\delta_{jk}E_{il}-\delta_{li}E_{kj}$. In other words, the 
elements $\{b_j^+,b_k^-\}$ form, up to a factor 2, the standard $\u(n)$ or $\gl(n)$
basis elements.

So the odd generators $b_j^\pm$ clearly reveal the subalgebra chain $\osp(1|2n) \supset \sp(2n) \supset \u(n)$.
Note that $\u(n)$ is, algebraically, the same as the general linear Lie algebra $\gl(n)$. 
But the condition $(b_j^\pm)^\dagger=b_j^\mp$ implies that we are dealing here with the ``compact form'' $\u(n)$.

The subalgebra $\u(n)$ can be extended to a parabolic subalgebra ${\cal P}$ of $\osp(1|2n)$~\cite{Palev4}:
\begin{equation}
{\cal P} = \hbox{span} \{ \{b_j^+, b_k^-\}, b_j^-, \{b_j^-, b_k^-\} \;|\;
j,k=1,\ldots,n \}.
\label{P}
\end{equation}

Recall that $\{b_j^-,b_k^+\} |0\rangle = p\,\delta_{jk}\, |0\rangle$, with $\{b_j^-,b_j^+\}=2 h_j$.
This means that the space spanned by $|0\rangle$ is a trivial one-dimensional $\u(n)$ module $\C |0\rangle$
of weight $(\frac{p}{2}, \ldots, \frac{p}{2})$.
Since $b_j^- |0\rangle =0$,  the module $\C |0\rangle$ can be extended to a one-dimensional ${\cal P}$ module.
Now we are in a position to define the induced $\osp(1|2n)$ module $\overline V(p)$:
\begin{equation}
 \overline V(p) = \hbox{Ind}_{\cal P}^{\osp(1|2n)} \C|0\rangle.
 \label{defInd}
\end{equation}
This is an $\osp(1|2n)$ representation with lowest weight $(\frac{p}{2}, \ldots, \frac{p}{2})$.
By the Poincar\'e-Birkhoff-Witt theorem~\cite{Kac1, Palev4}, it is easy to give a basis for  $\overline V(p)$:
\begin{align}
& (b_1^+)^{k_1}\cdots (b_n^+)^{k_n} (\{b_1^+,b_2^+\})^{k_{12}}  (\{b_1^+,b_3^+\})^{k_{13}} \cdots 
(\{b_{n-1}^+,b_n^+\})^{k_{n-1,n}} |0\rangle, \label{Vpbasis}\\
& \qquad k_1,\ldots,k_n,k_{12},k_{13}\ldots,k_{n-1,n} \in \Z_+. \nn
\end{align}
This is all rather formal, and it sounds easy. The difficulty however comes from the fact that
in general $\overline V(p)$ is not a simple module (i.e.\ not an irreducible representation) of
$\osp(1|2n)$. Let $M(p)$ be the maximal nontrivial submodule of $\overline V(p)$. Then the
simple module (irreducible module), corresponding to the paraboson Fock space, is
\begin{equation}
V(p) = \overline V(p) / M(p).
\label{Vp}
\end{equation}
The purpose is now to determine the vectors belonging to $M(p)$, and hence to find 
the structure of $V(p)$. Furthermore, we want to find explicit matrix elements of
the $\osp(1|2n)$ generators in an appropriate basis of $V(p)$. 
As an illustrative example, we shall first treat the case of $\osp(1|4)$.

\setcounter{equation}{0}
\section{Paraboson Fock representations of $\osp(1|4)$} \label{sec:osp14}

We shall examine the induced module $\overline V(p)$ in the case $n=2$, with basis vectors
\begin{equation}
|k,l,m\rangle \equiv  (b_1^+)^{k}(b_2^+)^{l} (\{b_1^+,b_2^+\})^{m} |0\rangle
\qquad (k,l,m\in\Z_+).
\label{klm}
\end{equation}
Note that the weight of this vector is
\begin{equation}
(\frac{p}{2}, \frac{p}{2}) + k \delta_1+ l\delta_2 +m (\delta_1+\delta_2).
\label{wt-klm}
\end{equation}
The {\em level} of this vector is defined as $k+l+2m$.
The basis vectors $|k,l,m\rangle$ are easy to define, and the actions of the generators
$b_1^\pm$, $b_2^\pm$ on these vectors can be computed, using the triple relations.
For the positive root vectors this is easy:
\begin{align}
& b_1^+ |k,l,m\rangle = |k+1,l,m\rangle,\nn\\
& b_2^+ |2k,l,m\rangle = |2k,l+1,m\rangle,\nn\\
& b_2^+ |2k+1,l,m\rangle = |2k,l,m+1\rangle -|2k+1,l+1,m\rangle.
\label{b+klm}
\end{align}
For the negative root vectors this requires some tough computations, yielding:
\begin{align}
& b_1^- |2k,l,m\rangle = 2k |2k-1,l,m\rangle + 2m |2k,l+1,m-1\rangle,\nn\\
& b_1^- |2k+1,l,m\rangle = (p+2m+2k) |2k,l,m\rangle - 2m |2k+1,l+1,m-1\rangle,\nn\\
& b_2^- |2k,2l,m\rangle = 2l |2k,2l-1,m\rangle + 2m |2k+1,2l,m-1\rangle,\nn\\
& b_2^- |2k,2l+1,m\rangle = (p+2l) |2k,2l,m\rangle + 2m |2k+1,2l+1,m-1\rangle,\nn\\
& b_2^- |2k+1,2l,m\rangle = 2l |2k,2l-2,m+1\rangle - 2l|2k+1,2l-1,m\rangle \nn\\
& \qquad\qquad+ 2m |2k+2,2l,m-1\rangle,\nn\\
& b_2^- |2k+1,2l+1,m\rangle = 2l |2k,2l-1,m+1\rangle - (p+2l-2)|2k+1,2l,m\rangle \nn\\
& \qquad\qquad + 2m |2k+2,2l+1,m-1\rangle.
\label{b-lkm}
\end{align}

It is now possible to compute ``inner products'' of vectors $|k,l,m\rangle$,
using $\langle0|0\rangle=1$ and $(b_i^\pm)^\dagger = b_i^\mp$. 
Clearly, vectors of different weight have inner product zero.
Let us compute a number of the nonzero inner products.
At weight $(\frac{p}{2}, \frac{p}{2})$ there is one vector only, $|0,0,0\rangle = |0\rangle$,
with
\begin{equation}
\langle 0,0,0 | 0,0,0 \rangle =1.
\label{level0}
\end{equation}
At level 1 there is one vector of weight $(\frac{p}{2}+1, \frac{p}{2})$ and one of weight
$(\frac{p}{2}, \frac{p}{2}+1)$, with inner products respectively:
\begin{equation}
\langle 1,0,0|1,0,0\rangle=p, \qquad 
\langle 0,1,0|0,1,0\rangle=p.
\label{level1}
\end{equation}
At level 2 there is one vector of weight $(\frac{p}{2}+2, \frac{p}{2})$, one of weight
$(\frac{p}{2}, \frac{p}{2}+2)$, and two vectors of weight $(\frac{p}{2}+1, \frac{p}{2}+1)$.
The inner products are given by:
\begin{align}
& \langle 2,0,0|2,0,0\rangle= \langle 0,2,0|0,2,0\rangle=2p, \nn\\
& \langle 1,1,0|1,1,0\rangle=p^2,\ \langle 1,1,0|0,0,1\rangle=2p,\ \langle 0,0,1|0,0,1\rangle=4p.
\label{level2}
\end{align}
{}From~\eqref{level1} it follows already that $p$ should be a positive number, otherwise 
the inner product (bilinear form) is not positive definite. 
The matrix of inner products of the vectors of weight $(\frac{p}{2}+1, \frac{p}{2}+1)$
has determinant
\begin{equation}
\det \left(\begin{array}{cc} p^2 & 2p \\ 2p & 4p \end{array}\right) = 4p^2(p-1).
\label{detlevel2}
\end{equation}
So this matrix is positive definite only if $p>1$. 
Thus, for $p>1$ both vectors of weight $(\frac{p}{2}+1, \frac{p}{2}+1)$ belong to $V(p)$;
but for $p=1$ one vector ($2|1,1,0\rangle - |0,0,1\rangle$) belongs to $M(p)$ and 
the subspace of $V(p)$ of weight $(\frac{p}{2}+1, \frac{p}{2}+1)$ is one-dimensional.

One could continue this analysis level by level, but the computations become rather
complicated and in order to find a technique that works for arbitrary $n$ one should
find a better way of analysing $\overline V(p)$.
For this purpose, we shall construct a different basis for $\overline V(p)$.
This new basis is indicated by the character of $\overline V(p)$: this is a formal
infinite series of terms $\mu x_1^{j_1} x_2^{j_2}$, with $(j_1,j_2)$ a weight of $\overline V(p)$
and $\mu$ the dimension of this weight space.
So the vacuum vector $|0\rangle$ of $\overline V(p)$, of weight $(\frac{p}{2},\frac{p}{2})$, yields a 
term $x_1^{\frac{p}{2}} x_2^{\frac{p}{2}} = (x_1x_2)^{p/2}$
in the character $\ch \overline V(p)$.
Since the basis vectors are given by $(b_1^+)^{k}(b_2^+)^{l} (\{b_1^+,b_2^+\})^{m} |0\rangle$,
where $k,l,m \in\Z_+$, it follows that
\begin{equation}
 \ch \overline V(p) = \frac{(x_1x_2)^{p/2}}{(1-x_1)(1-x_2)(1-x_1x_2)}.
 \label{char-osp14}
\end{equation}
Such expressions have an interesting expansion in terms of Schur functions,
valid for general $n$.
 
\begin{prop}[Cauchy, Littlewood]
Let $x_1,\ldots,x_n$ be a set of $n$ variables. Then~\cite{Little}
\begin{equation}
\frac{1}{\prod_{i=1}^n(1-x_i)\prod_{1\leq j<k\leq n}(1-x_jx_k)} = \sum_{\lambda} s_\lambda (x_1,\ldots,x_n)
= \sum_{\lambda} s_\lambda (x) 
\label{Schur}
\end{equation}
In the right hand side, the sum is over all partitions $\lambda$ and $s_\lambda(x)$ is the Schur symmetric 
function{\rm ~\cite{Mac}}. 
\end{prop}
For $n$ variables, $s_\lambda(x)=0$ if the length $\ell(\lambda)$ is greater than $n$, so in practice the sum is
over all partitions of length less than or equal to $n$.
For example, for $n=2$ one has
\[
 \ch \overline V(p) = (x_1x_2)^{p/2}(1+s_{1}(x)+ s_{2}(x)+s_{1,1}(x)+ s_{3}(x)+s_{2,1}(x)+ \cdots).
\]
The characters of finite dimensional $\u(n)$ representations (here $\u(2)$)
are given by such Schur functions $s_\lambda(x)$.
Hence such expansions are useful since they yield the branching to $\u(n)$ of the $\osp(1|2n)$
representation $\overline V(p)$.
But for such finite dimensional $\u(n)$ representations labelled by a partition
$\lambda$, there is a known basis: the Gelfand-Zetlin basis (GZ)~\cite{GZ, Baird}.
We shall use the $\u(n)$ GZ basis vectors as our new basis for $\overline V(p)$. 
For our current example, $n=2$, the new basis vectors are thus given by:
\begin{equation}
|m) = \left| p; \begin{array}{l} m_{12},m_{22} \\ m_{11} \end{array} \right)\equiv 
\left| \begin{array}{l} m_{12},m_{22} \\ m_{11} \end{array} \right),
\label{m2}
\end{equation}
where $(m_{12},m_{22})$ is the partition $\lambda$ of length~2 (that is
$m_{12}$ and $m_{22}$ are integers with $m_{12} \geq m_{22} \geq 0$),
and the GZ pattern $m$ satisfies the betweenness condition
\[
m_{12}\geq m_{11}\geq m_{22}.
\]
As $p$ is supposed to be fixed, it will usually be dropped from the notation of
the vectors $|m)$, as in~\eqref{m2}. For such vectors,
$(m_{12},m_{22})$ is the $\u(2)$ representation label, and $m_{11}$ is the $\u(2)$
internal label of the vector $|m)$.
We assume that the action of the $\u(2)$ generators is as usual, thus the weight of
the vector $|m)$ is given by
\[
(\frac{p}{2},\frac{p}{2}) + (m_{11},m_{12}+m_{22}-m_{11}).
\]

In this new basis, we need to compute the action of $b_1^\pm$ and $b_2^\pm$
on the basis vectors $|m)$. In fact, it will be sufficient to compute only those
of $b_1^+$ and $b_2^+$ (as $b_j^- = (b_j^+)^\dagger$).
Then, the relation with the old basis $|k,l,m\rangle$ follows implicitly from
$|0,0,0\rangle = \left| \begin{array}{l} 0,0 \\ 0 \end{array} \right)$.

{}From the triple relations~\eqref{pboson}, it follows that
under the $\u(2)$ basis~\eqref{un}, the set $(b_1^+, b_2^+)$ forms a tensor of rank (1,0),
with weights $\delta_1$ and $\delta_2$, and GZ patterns 
\begin{equation}
b_1^+ \sim \left( \begin{array}{cc}1&0\\1& \end{array}\right),\qquad
b_2^+ \sim \left( \begin{array}{cc}1&0\\0& \end{array}\right).
\label{b1b2}
\end{equation}
Therefore,
\begin{equation}
 ( m' | b_1^+ | m ) = 
\left( \begin{array}{ll} m_{12} & m_{22} \\ m_{11} & \end{array}; 
\begin{array}{cc}1&0\\1& \end{array} \right| \left.
\begin{array}{ll} m_{12}' & m_{22}' \\ m_{11}' & \end{array} \right)
\times
( m_{12}',m_{22}' || b^+ || m_{12},m_{22} ).
\label{matrixel}
\end{equation}
The first factor in the right hand side is a classical $\u(2)$ Clebsch-Gordan coefficient (CGC)~\cite{Klimyk, Bied},
and the second factor is a {\em reduced matrix element}~\cite{Klimyk}.
The possible values of the patterns $m'$ are determined by the $\u(2)$ tensor product
$ (1,0)\otimes (m_{12},m_{22}) = (m_{12}+1,m_{22}) \oplus (m_{12},m_{22}+1)$, and by the
additivity property of the internal labels ($m'_{11}=m_{11}+1$ in the above expression).
So the only $\u(2)$ CGCs of relevance are given below, their values taken from~\cite[p.~385]{Klimyk}:
\begin{align}
&\left( \begin{array}{l} m_{12},m_{22} \\ m_{11} \end{array}; 
 \begin{array}{l}1, 0\\1 \end{array} \right| \left.
 \begin{array}{l} m_{12}+1 , m_{22} \\ m_{11}+1 \end{array} \right) =
 \sqrt{\frac{m_{11}-m_{22}+1}{m_{12}-m_{22}+1}}, \\
&\left( \begin{array}{l} m_{12},m_{22} \\ m_{11} \end{array}; 
 \begin{array}{l}1, 0\\1 \end{array} \right| \left.
 \begin{array}{l} m_{12}, m_{22}+1 \\ m_{11}+1 \end{array} \right) =
 -\sqrt{\frac{m_{12}-m_{11}}{m_{12}-m_{22}+1}},\\
&\left( \begin{array}{l} m_{12},m_{22} \\ m_{11} \end{array}; 
 \begin{array}{l}1, 0\\0 \end{array} \right| \left.
 \begin{array}{l} m_{12}+1 , m_{22} \\ m_{11} \end{array} \right) =
 \sqrt{\frac{m_{12}-m_{11}+1}{m_{12}-m_{22}+1}},\\
&\left( \begin{array}{l} m_{12},m_{22} \\ m_{11} \end{array}; 
 \begin{array}{l}1, 0\\0 \end{array} \right| \left.
 \begin{array}{l} m_{12} , m_{22}+1 \\ m_{11} \end{array} \right) =
 \sqrt{\frac{m_{11}-m_{22}}{m_{12}-m_{22}+1}}.
\end{align} 
The problem is thus reduced to finding explicit expressions for two functions 
$F_1$ and $F_2$, where
\begin{equation}
F_1(m)=( m_{12}+1,m_{22} || b^+ || m_{12},m_{22} ), \qquad
F_2(m)=( m_{12},m_{22}+1 || b^+ || m_{12},m_{22} ).
\label{F1F2}
\end{equation}
We can write:
\begin{align}
b_1^+ | m ) & = \sqrt{\frac{m_{11}-m_{22}+1}{m_{12}-m_{22}+1}} F_1(m) 
\left| \begin{array}{l} m_{12}+1, m_{22} \\ m_{11}+1 \end{array} \right)
 -\sqrt{\frac{m_{12}-m_{11}}{m_{12}-m_{22}+1}} F_2(m) 
\left| \begin{array}{l} m_{12},m_{22}+1 \\ m_{11}+1 \end{array} \right),\label{b1+}\\
b_2^+ | m ) & =  \sqrt{\frac{m_{12}-m_{11}+1}{m_{12}-m_{22}+1}} F_1(m) 
\left| \begin{array}{l} m_{12}+1, m_{22} \\ m_{11} \end{array} \right)
+  \sqrt{\frac{m_{11}-m_{22}}{m_{12}-m_{22}+1}} F_2(m) 
\left| \begin{array}{l} m_{12},m_{22}+1 \\ m_{11} \end{array} \right), \label{b2+}
\end{align}
and the action of $b_j^-$ follows from $(m'|b_j^-|m) = (m|b_j^+|m')$.

Now it remains to determine the functions $F_1$ and $F_2$.
From the action 
\begin{equation}
\{ b_2^-, b_2^+ \} |m) = 2h_2 |m) = 
(p+2(m_{12}+m_{22}-m_{11})) |m), \label{b2b2}
\end{equation}
one deduces the following recurrence relations 
for $F_1$ and $F_2$:
\begin{align}
& \frac{F_1(m_{12},m_{22}) F_2(m_{12}+1,m_{22}-1)}{\sqrt{m_{12}-m_{22}+1}\sqrt{m_{12}-m_{22}+3}}
+ \frac{F_1(m_{12},m_{22}-1) F_2(m_{12},m_{22}-1)}{m_{12}-m_{22}+2} =0, \label{R1}\\
& \frac{m_{12}-m_{11}+1}{m_{12}-m_{22}+1}F_1(m_{12},m_{22})^2- \frac{m_{22}-m_{11}}{m_{12}-m_{22}+1}F_2(m_{12},m_{22})^2 
+\frac{m_{12}-m_{11}}{m_{12}-m_{22}}F_1(m_{12}-1,m_{22})^2\nn\\
&\qquad-\frac{m_{22}-m_{11}-1}{m_{12}-m_{22}+2}F_2(m_{12},m_{22}-1)^2 =
p+2m_{12}+2m_{22}-2m_{11}. \label{R2}
\end{align}
The action $b_2^-|m)$ leads to the boundary condition: $F_2(m_{12},m_{22}-1)=0$ if $m_{22}=0$.
The boundary condition together with the recurrence relations~\eqref{R1}-\eqref{R2}
lead to the following solution 
for the unknown functions $F_1$ and $F_2$:
\begin{align}
F_1(m_{12},m_{22}) & = (-1)^{m_{22}} (m_{12}+2+{\cal E}_{m_{12}}(p-2))^{1/2} 
\frac{(m_{12}-m_{22}+1)^{1/2}}{(m_{12}-m_{22}+1+{\cal O}_{m_{12}-m_{22}})^{1/2}}, \label{F1}\\
F_2(m_{12},m_{22}) &= (m_{22}+1+{\cal E}_{m_{22}}(p-2))^{1/2}
\frac{(m_{12}-m_{22}+1)^{1/2}}{(m_{12}-m_{22}+1-{\cal O}_{m_{12}-m_{22}})^{1/2}}, \label{F2}
\end{align}
where the {\em even} and {\em odd functions} ${\cal E}_j$ and ${\cal O}_j$ are defined by
\begin{align}
& {\cal E}_{j}=1 \hbox{ if } j \hbox{ is even and 0 otherwise},\nn\\
& {\cal O}_{j}=1 \hbox{ if } j \hbox{ is odd and 0 otherwise}. \label{EO}
\end{align}
The solution for $F_1$ and $F_2$ is unique up to a choice of the sign factor. 
At this moment, only the action of $\{b_2^-,b_2^+\}$ has been used in the process.
Now it remains to verify whether the actions of $b_1^\pm$ and $b_2^\pm$ thus determined
do indeed yield a solution, i.e.\ one should verify that all triple relations~\eqref{pboson}
are satisfied. This is a straightforward but tedious computation; the only result provided
by this calculation is that the sign factors are restricted. 
The choice of sign factors presented in~\eqref{F1}-\eqref{F2} is the simplest solution.

So we finally present the complete solution in the case of $\osp(1|4)$:
\begin{align}
b_1^+ | m )  =& \sqrt{m_{11}-m_{22}+1}\; f_1(m_{12},m_{22}) 
\left| \begin{array}{l} m_{12}+1, m_{22} \\ m_{11}+1 \end{array} \right) \nn\\
 & -\sqrt{m_{12}-m_{11}}\; f_2(m_{12},m_{22}) 
\left| \begin{array}{l} m_{12},m_{22}+1 \\ m_{11}+1 \end{array} \right),\label{sol1}\\
b_2^+ | m )  =& \sqrt{m_{12}-m_{11}+1}\; f_1(m_{12},m_{22}) 
\left| \begin{array}{l} m_{12}+1, m_{22} \\ m_{11} \end{array} \right) \nn\\
 & +\sqrt{m_{11}-m_{22}}\; f_2(m_{12},m_{22}) 
\left| \begin{array}{l} m_{12},m_{22}+1 \\ m_{11} \end{array} \right),\label{sol2}\\
b_1^- | m )  =& \sqrt{m_{11}-m_{22}}\; f_1(m_{12}-1,m_{22}) 
\left| \begin{array}{l} m_{12}-1, m_{22} \\ m_{11}-1 \end{array} \right) \nn\\
 & -\sqrt{m_{12}-m_{11}+1}\; f_2(m_{12},m_{22}-1) 
\left| \begin{array}{l} m_{12},m_{22}-1 \\ m_{11}-1 \end{array} \right),\label{sol3}\\
b_2^- | m )  =& \sqrt{m_{12}-m_{11}}\; f_1(m_{12}-1,m_{22}) 
\left| \begin{array}{l} m_{12}-1, m_{22} \\ m_{11} \end{array} \right) \nn\\
 & +\sqrt{m_{11}-m_{22}+1}\; f_2(m_{12},m_{22}-1) 
\left| \begin{array}{l} m_{12},m_{22}-1 \\ m_{11} \end{array} \right), \label{sol4}
\end{align}
where
\begin{align}
f_1(m_{12},m_{22}) &= (-1)^{m_{22}} \frac{(m_{12}+2+{\cal E}_{m_{12}}(p-2))^{1/2}}
{(m_{12}-m_{22}+1+{\cal O}_{m_{12}-m_{22}})^{1/2}}, \\
f_2(m_{12},m_{22}) &= \frac{(m_{22}+1+{\cal E}_{m_{22}}(p-2))^{1/2}}
{(m_{12}-m_{22}+1-{\cal O}_{m_{12}-m_{22}})^{1/2}}.
\end{align}

In this whole analysis we have assumed that the parameter $p$ is sufficiently large
such that all factors appearing under square root symbols are positive,
in other words such that $\overline V(p)$ itself is irreducible.
The general expressions thus obtained now allow us to examine the structure in
more detail. 
In the expressions~\eqref{F1}-\eqref{F2}, the first factor in the right hand side determines
the essential cases when the reduced matrix element is zero or not.
Let us depict these factors (or rather their squares) in a scheme 
as given in Figure~1.
In this diagram, there is an edge between two partitions $(m)$ and $(m')$ if the reduced
matrix element $(m'||b^+||m)$ is nonzero, and the number on top of the edge is
the crucial factor of $(m'||b^+||m)^2$.
Clearly, these factors tell us when these partitions ``are part of'' the irreducible
representation $V(p)$ or not (to be more precise, whether the vectors of the
$\u(2)$ module labelled by $(m)$ belong to $M(p)$ or not).
Obviously, for $p=1$ only the top line of the scheme survives, and forms the
$\u(2)$ content of $V(1)$. For $p>1$, all $\u(2)$ representations $(m)$ survive, 
and $V(p)=\overline V(p)$.

\begin{theo}
The $\osp(1|4)$ representation $V(p)$ with lowest weight $(\frac{p}{2},\frac{p}{2})$
is a unirrep if and only if $p\geq 1$.
For $p>1$,  $V(p)=\overline V(p)$ and 
\[
\ch V(p) = (x_1x_2)^{p/2} /((1-x_1)(1-x_2)(1-x_1x_2)).
\]
For $p=1$, $V(p)=\overline V(p)/M(p)$ with $M(p)\ne 0$. $V(1)$ is the boson Fock space, and
\[
\ch V(1) = (x_1x_2)^{1/2} /((1-x_1)(1-x_2)).
\]
\end{theo}
The explicit action of the $\osp(1|4)$ generators in $V(p)$ is given by~\eqref{sol1}-\eqref{sol4}. 
The basis is orthogonal and normalized. Note that also
for $p=1$ this action remains valid, provided one keeps in mind that all vectors with $m_{22}\ne 0$
must vanish.

We end this section mentioning that more general $\osp(1|4)$ irreducible representations with no unique
``vacuum" were investigated in~\cite{Heid, Blank}. However their matrix elements were not determined.
\setcounter{equation}{0}
\section{Paraboson Fock representations of $\osp(1|2n)$} \label{sec:Fock-osp12n}

Similarly as in the previous section, we start our analysis by considering the 
induced module $\overline V(p)$. First, a new basis for $\overline V(p)$ will be
introduced. In this basis, matrix elements are computed, and from these expressions it
will be clear which vectors belong to $M(p)$.

A basis for $\overline V(p)$ was already given in~\eqref{Vpbasis}. From this expression, one finds
\begin{equation}
\ch \overline V(p) = \frac{(x_1\cdots x_n)^{p/2}}{\prod_{i=1}^n(1-x_i)\prod_{1\leq j<k\leq n}(1-x_jx_k)}.
\label{char-osp12n}
\end{equation}
The denominator can be expanded as in~\eqref{Schur}. The relation between Schur functions
and $\u(n)$ characters~\cite{Weyl} makes it again natural to consider a basis consisting of 
all GZ-patterns~\cite{GZ, Baird}
for all possible partitions of length at most $n$. Thus the new basis of $\overline V(p)$
consists of vectors of the form
\begin{equation}
 |m)\equiv |m)^n\equiv \left|
\begin{array}{lcllll}
 m_{1n} & \cdots & \cdots & m_{n-1,n} & m_{nn}  \\
 m_{1,n-1} & \cdots & \cdots &  m_{n-1,n-1}  &  \\
\vdots & \qdots & & & \\
m_{11} & & & &
\end{array}
\right) 
= \left| \begin{array}{l} [m]^n \\[2mm] |m)^{n-1} \end{array} \right)
\label{mn}
\end{equation}
Just as for $\osp(1|4)$, the label $p$ is dropped in the notation of the vectors $|m)$.
Herein, the top line of the pattern, also denoted by the $n$-tuple $[m]^n$, is any
partition (consisting of non increasing nonnegative numbers). 
The remaining $n-1$ lines of the pattern will sometimes be denoted by $|m)^{n-1}$.
So all $m_{ij}$ in the above GZ-pattern are nonnegative integers,
satisfying the {\em betweenness conditions}
\begin{equation}
m_{i,j+1}\geq m_{ij}\geq m_{i+1,j+1}\qquad (1\leq i\leq j\leq n-1).
\label{between}
\end{equation}
Note that, since the weight of $|0\rangle$ is $(\frac{p}{2}, \ldots, \frac{p}{2})$, the weight
of the above vector is determined by
\begin{equation}
h_{k}|m)=\left(\frac{p}{2}+\sum_{j=1}^k m_{jk}-\sum_{j=1}^{k-1} m_{j,k-1}\right)|m).
\label{hkm}
\end{equation}

Now we use a technique similar as in the previous section.
By the triple relations, one obtains
\[
[\{ b_i^+, b_j^-\},b_k^+] = 2\delta_{jk} b_i^+.
\]
With the identification $\{ b_i^+, b_j^-\}=2E_{ij}$ in the standard $\u(n)$ basis,
this is equivalent to the action $E_{ij}\cdot e_k = \delta_{jk} e_i$. 
In other words, the triple relations imply that
$(b_1^+,b_2^+,\ldots,b_n^+)$ is a standard $\u(n)$ tensor of rank $(1,0,\ldots,0)$. 
This means that one can attach a unique GZ-pattern with top line $1 0 \cdots 0$ to every $b_j^+$,
corresponding to the weight $+\delta_j$. Explicitly:
\begin{equation}
b_j^+ \sim \begin{array}{l}1 0 \cdots 0 0 0\\[-1mm]
1 0 \cdots 0 0\\[-1mm] \cdots \\[-1mm] 0 \cdots 0\\[-1mm] \cdots\\[-1mm] 0 \end{array},
\label{bGZ}
\end{equation}
where the pattern consists of $j-1$ zero rows at the bottom, and the first $n-j+1$ rows are of the form
$1 0 \cdots 0$.
The tensor product rule in $\u(n)$ reads
\begin{equation}
([m]^n) \otimes (1 0\cdots 0) = ([m]^n_{+1}) \oplus ([m]^n_{+2}) \oplus \cdots \oplus([m]^n_{+n})
\label{untensor}
\end{equation}
where $([m]^n) = (m_{1n},m_{2n},\ldots, m_{nn})$ and a subscript $\pm k$ indicates an increment of 
the $k$th label by $\pm 1$:
\begin{equation}
([m]^n_{\pm k}) = (m_{1n},\ldots,m_{kn}\pm 1,\ldots, m_{nn}).
\label{m+k}
\end{equation}
In the right hand side of~\eqref{untensor}, only those components which are still partitions (i.e.\
consisting of nondecreasing integers) survive.

A general matrix element of $b_j^+$ can now be written as follows:
\begin{align}
( m' | b_j^+ | m ) & = 
\left( \begin{array}{ll} [m]^n_{+k} \\[1mm] |m')^{n-1} \end{array} \right| b_j^+
\left| \begin{array}{ll} [m]^n \\[1mm] |m)^{n-1} \end{array} \right) \nn\\
& = \left( \begin{array}{ll} [m]^n \\[2mm] |m)^{n-1} \end{array} ; \right.
 \begin{array}{l}1 0 \cdots 0 0\\[-1mm]
1 0 \cdots 0\\[-1mm]  \cdots\\[-1mm] 0 \end{array}  
\left| \begin{array}{ll} [m]^n_{+k} \\[2mm] |m')^{n-1} \end{array} \right)
\times
([m]^n_{+k}||b^+||[m]^n).
\label{mmatrix}
\end{align}
The first factor in the right hand side is a $\u(n)$ Clebsch-Gordan coefficient~\cite{Klimyk},
the second factor is a reduced matrix element.
By the tensor product rule, the first line of $|m')$ has to be of the form~\eqref{m+k},
i.e.\ $[m']^n = [m]^n_{+k}$ for some $k$-value.

The special $\u(n)$ CGCs appearing here are well known, and have fairly simple expressions.
They can be found, e.g.\ in~\cite{Klimyk}. They can be expressed by means of
{\em $\u(n)$-$\u(n-1)$ isoscalar factors} and $\u(n-1)$ CGC's, which on their turn are written by means
of $\u(n-1)$-$\u(n-2)$ isoscalar factors and $\u(n-2)$ CGC's, etc. 
The explicit form of the special $\u(n)$ CGCs appearing here is given in Appendix~A.
The actual problem is now converted into finding expressions for the reduced
matrix elements, i.e.\ for the functions $F_k([m]^n)$, for arbitrary
$n$-tuples of non increasing nonnegative integers $[m]^n=(m_{1n},m_{2n},\ldots,m_{nn})$:
\begin{equation}
F_k([m]^n) = F_k(m_{1n},m_{2n},\ldots,m_{nn}) = ([m]^n_{+k}||b^+||[m]^n).
\label{Fk}
\end{equation}
So one can write:
\begin{align}
b_j^+|m) & = \sum_{k,m'} \left( \begin{array}{ll} [m]^n \\[2mm] |m)^{n-1} \end{array}\right. ;
  \begin{array}{l}1 0 \cdots 0 0\\[-1mm]
1 0 \cdots 0\\[-1mm]  \cdots\\[-1mm] 0 \end{array} 
\left| \begin{array}{ll} [m]^n_{+k} \\[2mm] |m')^{n-1} \end{array} \right)
F_k([m]^n) \left|  \begin{array}{ll} [m]^n_{+k} \\[1mm] |m')^{n-1} \end{array} \right), \label{bj+n}\\
b_j^-|m) & = \sum_{k,m'} \left( \begin{array}{ll} [m]_{-k}^n \\[2mm] |m')^{n-1} \end{array}\right. ;
 \begin{array}{l}1 0 \cdots 0 0\\[-1mm]
1 0 \cdots 0\\[-1mm]  \cdots\\[-1mm] 0 \end{array}  
\left| \begin{array}{ll} [m]^n \\[2mm] |m)^{n-1} \end{array} \right)
F_k([m]_{-k}^n) \left|  \begin{array}{ll} [m]^n_{-k} \\[1mm] |m')^{n-1} \end{array} \right). \label{bj-n}
\end{align}
In order to determine the unknown functions $F_k$, one can again start from the following action:
\begin{equation}
\{ b_n^-, b_n^+ \} |m) = 2h_n |m) = 
(p+2( \sum_{j=1}^n m_{jn}-\sum_{j=1}^{n-1} m_{j,n-1} )) |m).
\label{bnbn}
\end{equation}
Expressing the left hand side by means of~\eqref{bj+n}-\eqref{bj-n}, using the explicit form of the 
CGCs and isoscalar factors given in Appendix~A (which are simple here since $j=n$), one finds a system
of coupled recurrence relations for the functions $F_k$.
Together with the appropriate boundary conditions, we have been able to solve this,
in particular using Maple. 
So our main computational result is:
\begin{prop}
\label{prop-main}
The reduced matrix elements $F_k$ appearing in the actions of $b_j^\pm$ on vectors
$|m)$ of $\overline V(p)$ are given by:
\begin{align}
F_{k}(m_{1n}, m_{2n},\ldots,  m_{nn})  = &
(-1)^{m_{k+1,n}+\cdots+m_{nn}} (m_{kn}+n+1-k+ {\cal E}_{m_{kn}}(p-n))^{1/2} \nn\\
& \times  \prod_{j\neq k=1}^{n} \left( \frac{m_{jn}-m_{kn}-j+k }{m_{jn}-m_{kn}-j+k-{\cal O}_{m_{jn}-m_{kn}} }
\right)^{1/2}, \label{Main}
\end{align}
where ${\cal E}$ and ${\cal O}$ are the even and odd functions defined in~\eqref{EO}.
\end{prop}
It would be unfeasible to present all the details of this computational result. 
Essentially, the proof consist of verifying that all triple relations~\eqref{pboson} hold
when acting on any vector $|m)$. Each such verification leads to an algebraic identity
in $n$ variables $m_{1n},\ldots, m_{nn}$.
In these computations, there are some intermediate verifications: e.g.\ the action
$\{b_j^+,b_k^-\} |m)$ should leave the top row of the GZ-patter $|m)$ invariant (since
$\{b_j^+,b_k^-\}$ belongs to $\u(n)$). Furthermore, it must yields (up to a factor 2) the known action of
the standard $\u(n)$ matrix elements $E_{jk}$ in the classical GZ-basis.

The purpose is now to deduce the structure of $V(p)$ from the general expression
of the matrix elements in $\overline V(p)$.
Just as for $\osp(1|4)$, this will be governed by the factor 
\[
(m_{kn}+n+1-k+ {\cal E}_{m_{kn}}(p-n))
\]
in the expression of $F_k([m]^n)$, since this is the only factor in the right hand side
of~\eqref{Main} that may become zero.
If this factor is zero or negative, the assigned vector $|m)$ belongs to $M(p)$.
The integers $m_{jn}$ satisfy $m_{1n}\geq m_{2n} \geq \cdots \geq m_{nn}\geq 0$. 
If $m_{kn}=0$ (its smallest possible value), then this factor in $F_k$ takes the value
$(p-k+1)$. So the $p$-values $1,2,\ldots,n-1$ will play a special role.
Let us again depict these factors in a scheme, shown here in Figure~2 for $n=3$.
In this diagram, there is an edge between two partitions $([m]^n)$ and $([m']^n)=([m]^n_{+k})$ if the 
reduced matrix element $([m]_{+k}^n||b^+||[m]^n)=F_k([m]^n)$ is in general nonzero.
For the boldface lines, the relevant factor $(m_{kn}+n+1-k+ {\cal E}_{m_{kn}}(p-n))$ is
equal to $p-1$; for the dotted lines, this factor is equal to $p-2$.
As a consequence, for $p=1$ the irreducible module $V(p)$ corresponds to the
first line of the scheme only, i.e.\ only partitions $([m]^n)$ of length~1 appear
(the length of a partition is the number of nonzero parts).
For $p=2$, the irreducible module $V(p)$ is composed of partitions $([m]^n)$ of length~1 or~2 only.
This observation holds in general, due to the factor $(p-k+1)$ for $F_k$.
This finally leads to the following result:

\begin{theo}
\label{theo-main}
The $\osp(1|2n)$ representation $V(p)$ with lowest weight $(\frac{p}{2},\ldots, \frac{p}{2})$
is a unirrep if and only if $p\in\{1,2,\ldots,n-1\}$ or $p>n-1$.\\
For $p>n-1$,  $V(p)=\overline V(p)$ and 
\begin{align}
\ch V(p) &= \frac{(x_1\cdots x_n)^{p/2}}{\prod_i(1-x_i) \prod_{j<k}(1-x_jx_k) } \label{charx}\\
&= (x_1\cdots x_n)^{p/2} \sum_\lambda s_\lambda (x) \label{chars}
\end{align}
For $p\in\{1,2,\ldots,n-1\}$, $V(p)=\overline V(p)/M(p)$ with $M(p)\ne 0$. The structure
of $V(p)$ is determined by
\begin{equation}
\ch V(p) = (x_1\cdots x_n)^{p/2} \sum_{\lambda,\ \ell(\lambda)\leq p}  s_\lambda (x) \label{charp}
\end{equation}
where $\ell(\lambda)$ is the {\em length} of the partition $\lambda$.
\end{theo}
The explicit action of the $\osp(1|2n)$ generators in $V(p)$ is given 
by~\eqref{bj+n}-\eqref{bj-n} or \eqref{Abj+}-\eqref{Abj-}, and the basis is
orthogonal and normalized. For $p\in\{1,2,\ldots,n-1\}$ this action remains valid, 
provided one keeps in mind that all vectors with $m_{p+1,n}\ne0$ must vanish.

Note that the first line of Theorem~\ref{theo-main} can also be deduced 
from~\cite{Dobrev}, where all lowest weight unirreps of $\osp(1|2n)$
are classified by means of their lowest weight.

There is an interesting question related to the character of $V(p)$. For $p>n-1$,
the character is written in the form~\eqref{charx}, i.e.\ with a genuine $\osp(1|2n)$
denominator (each factor in this denominator corresponds to a positive root $\alpha$
of $\osp(1|2n)$ for which $2\alpha$ is not a root).
Can the characters of $V(p)$ for $p\in\{1,2,\ldots,n-1\}$ be written in a
similar form? In other words, what is $E_p$ in the expression
\begin{equation}
\ch V(p) = \boldsymbol{x}^{p/2} \sum_{\lambda,\ \ell(\lambda)\leq p}  s_\lambda (x) 
= \boldsymbol{x}^{p/2}\frac{E_p}{\prod_i(1-x_i) \prod_{j<k}(1-x_jx_k) },
\label{charE}
\end{equation}
where we have used the short hand notation $\boldsymbol{x}=x_1\cdots x_n$.
Some initial computations lead us to the following conjecture:
\begin{equation}
E_p = \sum_{\eta} (-1)^{c_\eta} s_\eta (x),
\label{E}
\end{equation}
where the sum is over all partitions $\eta$ of the form
\begin{equation}
\eta = \left(
\begin{array}{cccc}
a_1 & a_2 & \cdots & a_r \\
a_1+p & a_2+p & \cdots & a_r+p
\end{array} \right)
\label{eta}
\end{equation}
in Frobenius notation, and 
\begin{equation}
c_\eta= a_1+a_2+\cdots+a_r + r.
\label{c_eta}
\end{equation}
The  notation is a special way of denoting partitions, see~\cite{Mac}, related to 
the lengths of rows and columns in the Young diagram of the partition, counted from the diagonal.
In the current case, the partitions $\eta$ are all those with a Young diagram of the shape 
depicted in Figure~3.

Since the number of variables $x_1,\ldots, x_n$ is finite, the expression $E_p$
is also finite. Some special cases are:
\begin{align}
E_1 & = \prod_{1\leq j<k\leq n} (1-x_jx_k), \label{E1}\\
E_{n-1} & = 1-x_1x_2\cdots x_n. \label{En-1}
\end{align}
The first expression is a known S-function series, also due to Littlewood~\cite{Little}.
The first special case leads to
\begin{equation}
\ch V(1) = \frac{\boldsymbol{x}^{1/2} }{\prod_i(1-x_i) }, \label{V1}
\end{equation}
as it should be, since this corresponds to the ordinary boson Fock space.
The second special case yields
\begin{equation}
\ch V(n-1) = \boldsymbol{x}^{p/2}\frac{(1-x_1x_2\cdots x_n)}{\prod_i(1-x_i) \prod_{j<k}(1-x_jx_k) }
\label{Vn-1}
\end{equation}

Our conjecture that
\begin{equation}
\sum_{\lambda,\ \ell(\lambda)\leq p}  s_\lambda (x) = 
\frac{ \sum_{\eta} (-1)^{c_\eta} s_\eta (x) }{\prod_i(1-x_i) \prod_{j<k}(1-x_jx_k) }
\label{conj}
\end{equation}
was proved by R.C.\ King~\cite{King}. In fact, he also obtained an alternative expression in terms of 
determinants, which will be presented below. It leads to the following
alternative forms of the character, for $p=1,2,\ldots,n$:
\begin{align}
\ch V(p) &= \boldsymbol{x}^{p/2} \sum_{\lambda,\ \ell(\lambda)\leq p}  s_\lambda (x) \nn\\
& = \boldsymbol{x}^{p/2}\frac{ \sum_{\eta} (-1)^{c_\eta} s_\eta (x) }{\prod_i(1-x_i) \prod_{j<k}(1-x_jx_k) } \nn \\
& = \boldsymbol{x}^{p/2} \frac  {\det\left( x_i^{n-j}-(-1)^p\chi(j>p)x_i^{n-p+j-1} \right)_{i,j=1}^n }
  { \det \left(x_i^{n-j}-x_i^{n+j-1} \right)_{i,j=1}^n } \nn
\end{align}
where  $\chi(w)=1$ if $w$ is true and $0$ otherwise.

\setcounter{equation}{0}
\section{Branching to $\sp(2n)$ and $\sp(2n)$ characters} \label{sec:sp2n}

The $\osp(1|2n)$ representations $V(p)$ have been completely determined,
including the action of the $\osp(1|2n)$ generators $b_j^\pm$.
A basis for the even subalgebra $\sp(2n)$ has been given in~\eqref{sp2n}.
As a consequence, the explicit action of the $\sp(2n)$ basis elements
$\{ b_j^\pm, b_k^\pm\}$ can also be determined.
Under the action of $\sp(2n)$, however, $V(p)$ is in general not irreducible.
In this section we shall determine the decomposition of $V(p)$ in
irreducible $\sp(2n)$ components. We shall first consider the generic
case $p>n-1$, where $V(p)=\overline V(p)$.

Since $V(p)$ is a lowest weight representation, all irreducible $\sp(2n)$ components
will also be lowest weight representations. Each such component is thus characterized
by a lowest weight vector $v$. Such a vector should satisfy:
\begin{equation}
\{ b_j^-, b_k^-\}\cdot v =0 \hbox{ for } 1\leq j,k\leq n,\qquad
\{ b_j^-, b_k^+\}\cdot v =0 \hbox{ for } 1\leq j<k\leq n.
\label{v}
\end{equation}
From the explicit action of $b_j^\pm$ it is possible to deduce the following
\begin{prop}
The $n+1$ vectors $v_k$ ($k=0,1,\ldots,n$) of the form
\begin{equation}
v_k = \left| 
 \begin{array}{l}
 1 1 \cdots 1 1 0 \cdots 0 \\
 1 1 \cdots 1 0 \cdots 0 \\
 \cdots\\
 1 0 \cdots 0\\
 0 \cdots 0 \\
 \cdots\\
 0
 \end{array} \right) \qquad
 \begin{array}{l}
 \hbox{row }1:\ k \hbox{ ones} \\
 \hbox{row }2:\ k-1 \hbox{ ones} \\
 \cdots\\
 \hbox{row }k:\ 1 \hbox{ one}\\
 \hbox{row }k+1: \hbox{ all zeros} \\
 \cdots\\
 \hbox{row }n: \hbox{ zero}
 \end{array}
 \label{vk}
\end{equation}
satisfy~\eqref{v}. These are the only vectors of $V(p)$ satisfying~\eqref{v}.
The weight of $v_k$ is given by
\begin{equation}
(\frac{p}{2},\cdots, \frac{p}{2})+ (0,\ldots,0,1,\ldots,1) \qquad (k\hbox{ ones}).
\label{wt-vk}
\end{equation}
\end{prop}
As a consequence, the decomposition of $V(p)$ with respect to $\sp(2n)$ can be
written as
\begin{equation}
V(p) \rightarrow \bigoplus_{k=0}^n V(p,k)
\label{branching}
\end{equation}
where $V(p,k)$ is a unirrep of $\sp(2n)$ with lowest weight vector $v_k$
and lowest weight given by~\eqref{wt-vk}.

The structure of these unirreps $V(p,k)$ can be deduced from the character formula
of $V(p)$. Since 
\[
\prod_{i=1}^n (1+x_i) = 1 + s_1(x)+s_{1,1}(x)+\cdots + s_{1,1,\ldots,1}(x)
\]
it follows that
\begin{equation}
\ch V(p) = \frac{ \boldsymbol{x}^{p/2} }{\prod_i(1-x_i) \prod_{j<k}(1-x_jx_k) } =
\frac{ \boldsymbol{x}^{p/2}( 1 + s_1(x)+s_{1,1}(x)+\cdots + s_{1,1,\ldots,1}(x)) }
{\prod_i(1-x_i^2) \prod_{j<k}(1-x_jx_k) } .
\label{char2}
\end{equation}
Using the knowledge of the $\sp(2n)$ lowest weight vectors, one finds
\begin{prop}
The characters of the $\sp(2n)$ unirreps $V(p,k)$ with lowest weight~\eqref{wt-vk} and
with $p>n-1$ are given by
\begin{equation}
\ch V(p,k) = \boldsymbol{x}^{p/2} \frac{s_{1,\ldots,1}(x)}{\prod_i(1-x_i^2) \prod_{j<k}(1-x_jx_k) }
\label{charsp}
\end{equation}
where $(1,\ldots,1)=1^k$.
\end{prop}

For the non-generic case, $V(p)$ with $p\in\{1,2,\ldots,n-1\}$, the analysis is similar
and we will not give all the details. In fact, the vectors $v_k$ with $k=0,1,\ldots,p$
are again $\sp(2n)$ lowest weight vectors (this time $k\leq p$ because otherwise $v_k$
would belong to $M(p)$). This yields:
\begin{equation}
V(p) \rightarrow \bigoplus_{k=0}^p V(p,k).
\label{branchingp}
\end{equation}
The characters are essentially the same as in the generic case, except that one has
to exclude $\u(n)$ components with partitions of length greater than~$p$ (as the
corresponding reduced matrix elements vanish). So we can write
\begin{equation}
\ch V(p,k) = \boldsymbol{x}^{p/2} {\cal L}_p\left( 
\frac{s_{1,\ldots,1}(x)}{\prod_i(1-x_i^2) \prod_{j<k}(1-x_jx_k) } \right)
\label{Lp}
\end{equation}
where ${\cal L}_p$ indicates that in the Schur function expansion of its
argument one should keep only those $s_\lambda(x)$ with $\ell(\lambda)\leq p$.

Note that not all these $\sp(2n)$ characters are new. 
Holomorphic discrete series representations and harmonic series representations,
with characters of the type~\eqref{charsp}, have been determined in~\cite{King1}. 
In~\cite{Rowe}, characters of the type~\eqref{Lp} appear. 
Here, however, we have deduced not only
the character but also the explicit matrix elements.

\setcounter{equation}{0}
\section{Conclusions and remarks} \label{sec:conclusions}

In this paper, we have given a solution to a problem that has been open for many years,
namely giving the explicit structure of paraboson Fock representations. 
In order to solve this problem, we have used a combination of known techniques and
new computational power. We used in particular: the relation with unirreps of the 
Lie superalgebra $\osp(1|2n)$, the decomposition of the induced module $\overline V(p)$
with respect of the compact subalgebra $\u(n)$, the known GZ-basis for $\u(n)$ representations,
the method of reduced matrix elements for $\u(n)$ tensor operators and known
expressions for certain $u(n)$ CGCs and isoscalar factors. 

The solution given here is also the explicit solution that would be obtained 
by means of the Green ansatz~\cite{Green}. The method of Green's ansatz is easy to describe,
but difficult to perform, and has not lead to the explicit solution
of the paraboson Fock representations, as presented here.
In representation theoretic terms, Green's ansatz amounts to considering the $p$-fold
tensor product of the boson Fock space, $V(1)^{\otimes p}$, and extracting in this tensor
product the irreducible component with lowest weight $(\frac{p}{2},\ldots, \frac{p}{2})$.  

{}From the point of view of quantum field theory, one is often interested in the case that
$n=+\infty$. It is not easy to see what happens with the explicit matrix elements in that case.
On the other hand, the character formulas~\eqref{charp} and~\eqref{charE} are easy to describe under
this limit $n\rightarrow +\infty$, as Schur functions of an infinite number of variables are
common objects. 

Note that in Theorem~\ref{theo-main} in the second case, $p\in\{1,2,\ldots,n-1\}$, the
induced module $\overline V(p)$ is not irreducible. So it must have certain primitive 
vectors with respect to $\osp(1|2n)$. In fact, the weights of these primitive vectors
follow from~\eqref{charE} and~\eqref{E}.

In this paper we have constructed $\osp(1|2n)$ unirreps with a particular lowest weight,
namely of the form $(\frac{p}{2},\ldots, \frac{p}{2})$. It looks as if our techniques could
also be used to construct explicitly $\osp(1|2n)$ unirreps with 
a more general lowest weight of the form $(p_1,\ldots, p_n)$, though the
computational difficulties might be extremely hard.

\appendix 
\section{Appendix}
\renewcommand{\theequation}{\Alph{section}.\arabic{equation}}
\setcounter{equation}{0}

In this appendix we shall give the special $\u(n)$ CGCs~\cite{Klimyk, Bied} needed in the action~\eqref{bj+n}-\eqref{bj-n}
of the $\osp(1|2n)$ generators.
For this purpose, let us write $[m]^j$ for the $j$th row in a GZ-pattern, and we shall denote
the increment of element $i$ in such a $j$-tuple by $[m]^j_{\pm i}$:
\begin{align}
& [m]^{j}=[m_{1j}, m_{2j},\ldots , m_{jj}], \\
& [m]_{\pm i}^j=[m_{1j}, \ldots , m_{ij} \pm 1, \ldots, m_{jj}].
\end{align}
The $\u(n)$ CGCs can now be written as (a $\dot{0}$ stands for a sequence of zeros of appropriate length)
\begin{align}
& \left(
\begin{array}{l} [m]^n  \\ {[m]}^{n-1} \\ \cdots \\ {[m]^j} \\ {[m]}^{j-1}\\ \cdots \\ m_{11} \end{array} \right. ;
\begin{array}{l} 1\dot{0}\\ 1\dot{0} \\ \cdots \\ 1\dot{0}\\ \dot{0} \\ \cdots \\ 0 \end{array} 
\left| 
\begin{array}{l} [m]^n_{+k}  \\ {[m']}^{n-1} \\ \cdots \\ 
 {[m']^j} \\ {[m']}^{j-1}\\ \cdots \\ m'_{11} \end{array} \right)  \nn\\
& = \sum_{m'}
\left( \begin{array}{l}  [m]^n \\ {[m]}^{n-1} \end{array} \right.
\left| \begin{array}{l} 1\dot{0} \\ \epsilon \dot{0} \end{array} \right|
\left. \begin{array}{l} [m]^n_{+k}  \\ {[m']}^{n-1} \end{array} \right) \times
\left(
\begin{array}{l} {[m]}^{n-1} \\ \cdots \\ {[m]^j} \\ {[m]}^{j-1}\\ \cdots \\ m_{11} \end{array} \right. ;
\begin{array}{l} 1\dot{0} \\ \cdots \\ 1\dot{0}\\ \dot{0} \\ \cdots \\ 0 \end{array} 
\left| 
\begin{array}{l} {[m']}^{n-1} \\ \cdots \\ 
 {[m']^j} \\ {[m']}^{j-1}\\ \cdots \\ m'_{11} \end{array} \right) . 
\end{align}
In the right hand side, the first factor is an isoscalar factor, and the second factor
is a CGC of $\u(n-1)$.
The middle pattern in the $\u(n-1)$ CGC is that of the $\u(n)$ CGC with the first row deleted.
The middle pattern in the isoscalar factor consists of the first two rows of the middle pattern in the
left hand side, so $\epsilon$ is 0 or 1. 
If $\epsilon=0$, then $[m']^{n-1}=[m]^{n-1}$.
If $\epsilon=1$ then $[m']^{n-1}=[m_{1,n-1},\ldots, m_{r,n-1}+1,\ldots,m_{n-1,n-1}]=[m]^{n-1}_{+r}$
for some $r$-value.

The explicit forms of these isoscalar factors is given in~\cite[p.~385]{Klimyk}. We have:
\begin{equation}
\left( \begin{array}{l} [m]^n \\ {[m]}^{n-1} \end{array} \right.
\left| \begin{array}{l} 1 \dot{0} \\0 \dot{0} \end{array} \right|
\left. \begin{array}{l} [m]^n_{+k} \\ {[m]}^{n-1} \end{array} \right)
= \left( \frac{\prod_{j=1}^{n-1}  
(l_{j,n-1}-l_{kn}-1 )}
{ \prod_{j\neq k=1}^{n} (l_{jn}-l_{kn})}
\right)^{1/2} 
\label{iso1}
\end{equation}
where, as usual in this context, 
\begin{equation}
l_{ij}=m_{ij}-i;
\label{lij}
\end{equation}
and
\begin{equation}
\left( \begin{array}{l} [m]^n \\ {[m]}^{n-1} \end{array} \right.
\left| \begin{array}{l} 1 \dot{0} \\ 1 \dot{0} \end{array} \right|
\left. \begin{array}{l} [m]^n_{+k} \\ {[m]}^{n-1}_{+r} \end{array} \right)
= S(k,r)\left( \frac{{\prod_{j\neq r=1}^{n-1}}  
(l_{j,n-1}-l_{kn}-1 )\prod_{j\neq k=1}^{n}(l_{jn}-l_{r,n-1})}
{ \prod_{j\neq k=1}^{n} (l_{jn}-l_{kn})\prod_{j\neq r=1}^{n-1}(l_{j,n-1}
 -l_{r,n-1}-1)}
\right)^{1/2},
\label{iso2}
\end{equation}
where
\begin{equation}
S(k,r) = \left\{ \begin{array}{rcl}
 {1} & \hbox{for} & k\leq r  \\ 
 {-1} & \hbox{for} & k>r .
 \end{array}\right.
 \label{S}
\end{equation} 

We are now in a position to write down the explicit actions of the $\osp(1|2n)$ generators
on the basis vectors $|m)$. First, note that for the diagonal elements one has
\begin{equation}
h_{k}|m)=\left(\frac{p}{2}+\sum_{j=1}^k m_{jk}-\sum_{j=1}^{k-1} m_{j,k-1}\right)|m), 
\qquad (1\leq k\leq n). \label{h_k}
\end{equation}
Due to the simplicity of the CGCs, the actions of $b_n^\pm$ are the simplest to describe:
\begin{align}
b_{n}^+|m)&= \sum_{i=1}^n \left( \frac{\prod_{k=1}^{n-1}  
(l_{k,n-1}-l_{in}-1 )} { \prod_{k\neq i=1}^{n} (l_{kn}-l_{in})}
\right)^{1/2} F_{i}(m_{1n}, m_{2n},\ldots,  m_{nn}) \; |m)_{+in}; \label{bnp} \\
b_{n}^-|m)&= \sum_{i=1}^n \left( \frac{\prod_{k=1}^{n-1}  
(l_{k,n-1}-l_{in} )} { \prod_{k\neq i=1}^{n} (l_{kn}-l_{in}+1)}
\right)^{1/2} F_{i}(m_{1n},\ldots, m_{in}-1,\ldots,  m_{nn}) \; |m)_{-in}. \label{bnm} 
\end{align}
Herein, $F_i$ are the functions given in~\eqref{Main}, and $|m)_{\pm in}$ indicates the replacement
$m_{in}\rightarrow m_{in}\pm 1$.

Finally, the actions of the remaining generators $b_j^\pm$ ($j=1,2,\ldots,n-1$) are somewhat more involved,
due to the various isoscalar factors. They can be written in the following form:
\begin{align}
b_{j}^+|m)&=\sum_{i_n=1}^n\sum_{i_{n-1}=1}^{n-1}\ldots \sum_{i_j=1}^j
S(i_n,i_{n-1}) S(i_{n-1},i_{n-2})\ldots S(i_{j+1},i_{j})
\left( \frac{\prod_{k= 1}^{j-1}  
(l_{k,j-1}-l_{i_{j},j}-1 ) }
{ \prod_{k\neq i_{j}=1}^{j} (l_{kj}-l_{i_{j},j})}
\right)^{1/2} 
  \nn\\
&\times \prod_{r=1}^{n-j}
\left( \frac{\prod_{k\neq i_{n-r}=1}^{n-r}  
(l_{k,n-r}-l_{i_{n-r+1},n-r+1}-1 )\prod_{k\neq i_{n-r+1}=1}^{n-r+1}(l_{k,n-r+1}-l_{i_{n-r},n-r})
 }{ \prod_{k\neq i_{n-r+1}=1}^{n-r+1} (l_{k,n-r+1}-l_{i_{n-r+1},n-r+1})\prod_{k\neq i_{n-r}=1}^{n-r}(l_{k,n-r}
 -l_{i_{n-r},n-r}-1)}
\right)^{1/2}  \nn \\
&\times F_{i_n}(m_{1n},m_{2n},\ldots,  m_{nn}) \;
|m)_{+i_n,n;+i_{n-1},n-1;\ldots ;+i_j,j} ; \label{Abj+} \\[2mm]
b_{j}^-|m)&=\sum_{i_n=1}^n\sum_{i_{n-1}=1}^{n-1}\ldots \sum_{i_j=1}^j
S(i_n,i_{n-1}) S(i_{n-1},i_{n-2})\ldots S(i_{j+1},i_{j}) 
\left( \frac{\prod_{k= 1}^{j-1}  
(l_{k,j-1}-l_{i_{j},j} ) }
{ \prod_{k\neq i_{j}=1}^{j} (l_{kj}-l_{i_{j},j}+1)}
\right)^{1/2} 
 \nn\\
&\times \prod_{r=1}^{n-j}
\left( \frac{\prod_{k\neq i_{n-r}=1}^{n-r}  
(l_{k,n-r}-l_{i_{n-r+1},n-r+1} )\prod_{k\neq i_{n-r+1}=1}^{n-r+1}(l_{k,n-r+1}-l_{i_{n-r},n-r}+1)
 }{ \prod_{k\neq i_{n-r+1}=1}^{n-r+1} (l_{k,n-r+1}-l_{i_{n-r+1},n-r+1}+1)\prod_{k\neq i_{n-r}=1}^{n-r}(l_{k,n-r}
 -l_{i_{n-r},n-r})}
\right)^{1/2} \nn \\
&\times  F_{i_n}(m_{1n},\ldots,m_{i_n,n}-1,\ldots,  m_{nn}) \;
|m)_{-i_n,n;-i_{n-1},n-1;\ldots ;-i_j,j} . \label{Abj-}
\end{align}
Herein, each symbol $\pm i_k,k$ attached as a subscript to $|m)$ indicates a
replacement $m_{i_k,k}\rightarrow m_{i_k,k}\pm 1$. 

\section*{Acknowledgments}
The authors would like to thank Professor T.D.~Palev for his interest. 
NIS was supported by a project from the Fund for Scientific Research -- Flanders (Belgium)
and by project P6/02 of the Interuniversity Attraction Poles Programme (Belgian State -- 
Belgian Science Policy).

\newpage
\ 
\begin{figure}[t]
\caption{Relevant factors of the reduced matrix elements between two partitions (case of $\osp(1|4)$).}
\begin{center}
	\includegraphics[scale=0.8]{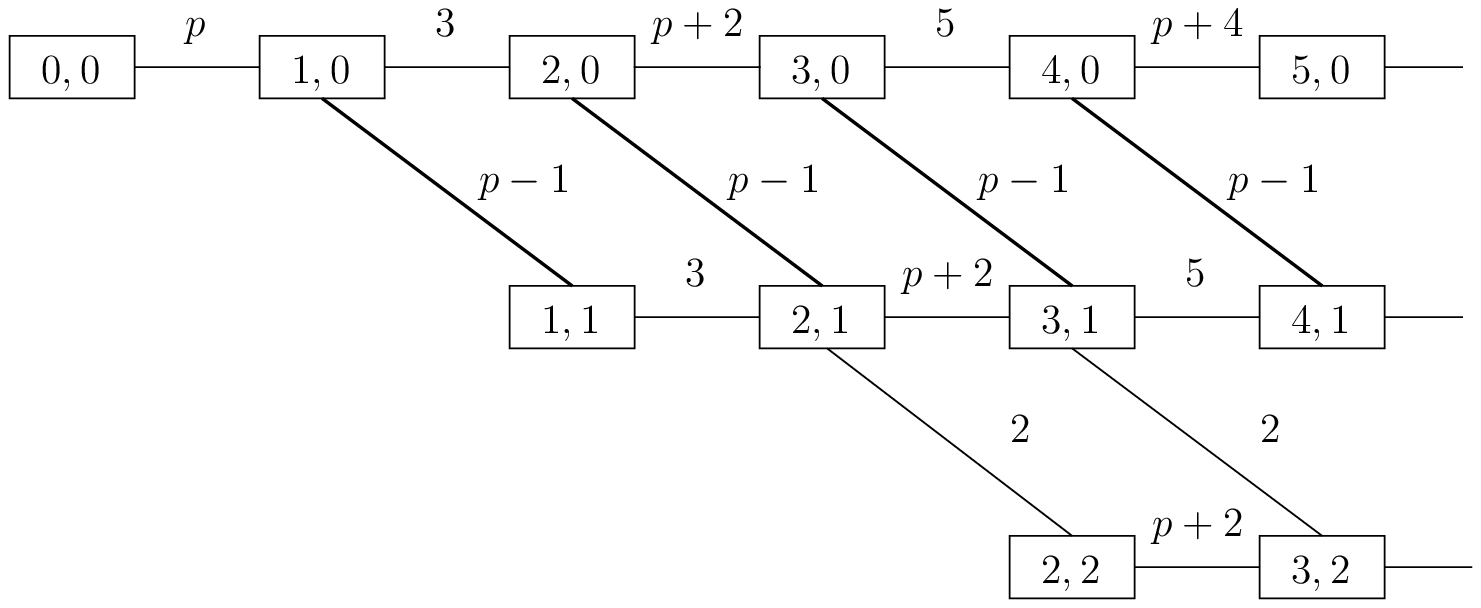}
\end{center}	
\end{figure}
\ 
\vspace{10cm}

\newpage
\ 
\begin{figure}[t]
\caption{Structure of the $\osp(1|2n)$ representations $\overline V(p)$ and $V(p)$, 
here illustrated for $n=3$ (so only partitions
of length at most~3 appear). The boldface line indicates that the corresponding reduced matrixelement has
a factor $(p-1)$; the dotted line indicates that it has a factor $(p-2)$.}
\begin{center}
\includegraphics[scale=0.9]{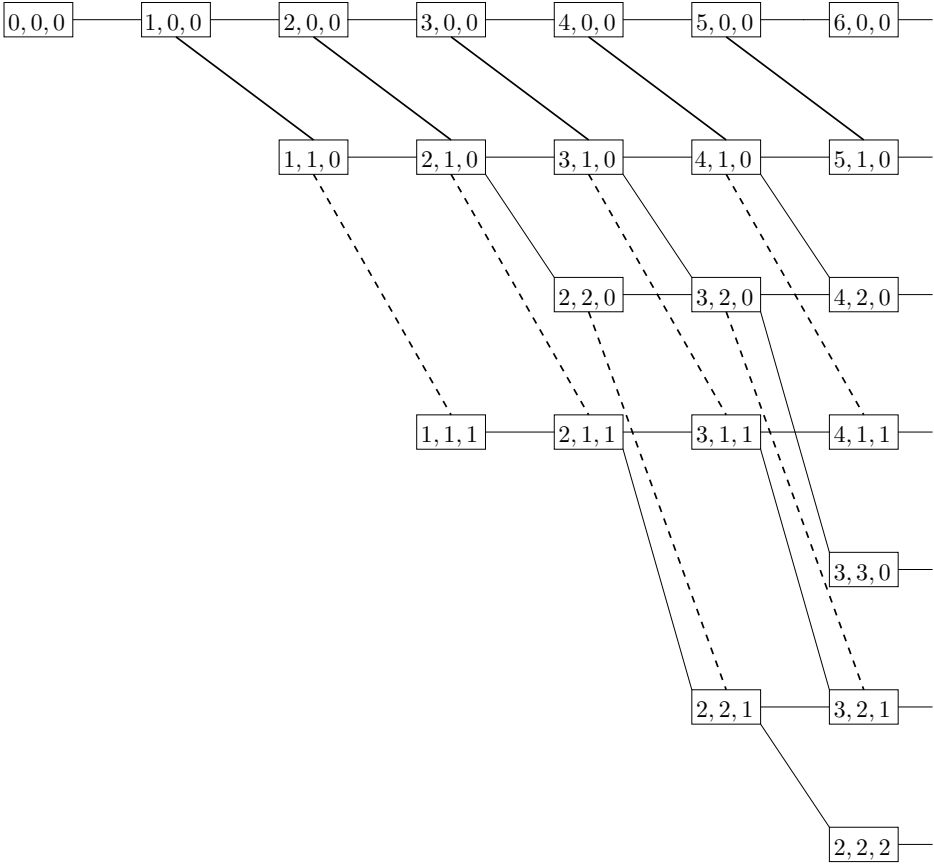}
\end{center}
\end{figure}
\ 
\vspace{10cm}

\newpage
\ 
\begin{figure}[t]
\caption{Typical shape of the Young diagram for the partition $\eta$, given by the Frobenius 
notation~\eqref{eta} (illustrated here for $r=3$).}
\begin{center}
\includegraphics[scale=0.7]{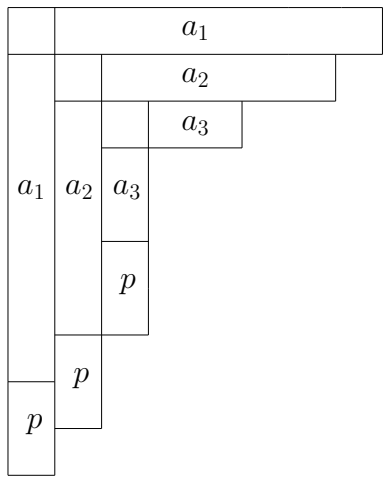}
\end{center}
\end{figure}
\ 
\vspace{10cm}

\end{document}